\documentclass[preprint,12pt]{elsarticle}
\usepackage{amssymb}

\journal{Computer Physics Communications}

\begin{document}

\begin{frontmatter}

\title{PolyMorph: Extension of PolyHoop for tissue morphogenesis coupled to chemical signaling}
\date{March 5, 2025}

\author[a]{Nicolas Pascal Guido M\"{u}ller}
\author[a,b]{Roman Vetter\corref{author}}

\cortext[author] {Corresponding author.\\\textit{E-mail address:} vetterro@ethz.ch}
\address[a]{Department of Biosystems Science and Engineering, ETH Z\"{u}rich, Schanzenstrasse 44, 4056 Basel, Switzerland}
\address[b]{Swiss Institute of Bioinformatics, Schanzenstrasse 44, 4056 Basel, Switzerland}

\begin{abstract}
We present PolyMorph, a lightweight standalone C++ program that extends its predecessor PolyHoop by a finite-difference solver for multi-component reaction-advection-diffusion equations. PolyMorph simulates two integral parts of tissue morphogenesis in two dimensions: 1) the mechanics of cellular deformation, growth and proliferation, and 2) transport and reaction of an arbitrary number of chemical species. Both of these components are bidirectionally coupled, allowing cells to base their behavior on local information on concentrations and flow, and allowing the chemical transport and reaction kinetics to depend on spatial information such as the local cell type. This bidirectional feedback makes PolyMorph a versatile tool to study a variety of cellular morphogenetic processes such as chemotaxis, cell sorting, tissue patterning with morphogen gradients, Turing patterning, and diffusion- or supply-limited growth with sub-cellular resolution.\\

\noindent {\bf NEW VERSION PROGRAM SUMMARY}

\begin{small}
	\noindent
	{\em Program Title:} PolyMorph \\
	{\em CPC Library link to program files:} TBA \\
	{\em Licensing provisions:} BSD 3-clause \\
	{\em Programming language:} C++11\\
	{\em Supplementary material:} Figure \ref{fig:1} \\
	{\em Journal reference of previous version:} Comput.\ Phys.\ Commun.\ 299 (2024) 109128, https://doi.org/10.1016/j.cpc.2024.109128 \\
	{\em Does the new version supersede the previous version?:} No \\[2ex]
	{\em Nature of problem:} In tissue development and disease, morphogenesis and cell fate determination depends on mechanical processes as well as chemical signaling. PolyMorph couples the Newtonian mechanics of deformable cells (including growth and proliferation) in 2D with a customizable set of reaction-advection-diffusion equations to simulate problems that require an integrated approach with chemical-mechanical interactions. Typical use cases include the patterning of epithelial tissues with chemical signals (e.g., morphogen gradients or the Turing mechanism), chemotaxis and cell migration, wound healing, diffusion- or nutrition-limited growth, regulatory network dynamics in a spatial cellular environment, and other problems in tissue self-organization. PolyMorph enables the numerical solution of such problems with bidirectional feedback between mechanics and chemistry, in large monolayer tissues and with an arbitrary number of interacting species.\\[2ex]
	{\em Solution method:} The off-lattice polygonal representation of cell boundaries in PolyHoop [1] is coupled to a lattice representation of diffusing chemical reactants. The reaction-advection-diffusion problem is solved with the finite difference method using the standard 5-point central difference stencil, and explicitly integrated in time. A scatter-gather approach inspired by the particle-in-cell method interpolates between the Lagrangian cell boundaries and the Eulerian finite-difference grid. The program is parallelized with OpenMP and does not use any external libraries.\\[2ex]
	{\em Reasons for the new version:} The original program PolyHoop solves the Newtonian mechanics of deformable particles, foams and cellular tissues in two dimensions. In the application realm of developmental and systems biology, however, various morphogenetic problems depend not only on cell mechanics, but also on chemical signaling, nutrient supply etc. The solution of such external transport problems was not part of PolyHoop, but is available in some related codes [2--5]. We have extended PolyHoop to create a new program named PolyMorph, introduced here. It enables the simulation of cellular tissue dynamics with bidirectional coupling to a multi-component diffusion system, which substantially widens the potential scope of application.\\[2ex]
	{\em Summary of revisions:} PolyMorph introduces a solver for $n$ coupled reaction-advection-diffusion equations for concentrations $c_i$, of the form $$\frac{\partial c_i}{\partial t}=D_i\left(\frac{\partial^2}{\partial x^2}+\alpha_i\frac{\partial^2}{\partial y^2}\right)c_i-\nabla\cdot(\vec{v}c_i)+R_i(p,c_1,...,c_n,k_1,...,k_m,t).$$ The reaction terms $R_i$ can be user-specified in the form of C++11 lambda functions, parameterized by the local polygon (cell) $p$ and a set of $m$ kinetic coefficients $k_1,...,k_m$. Both the kinetic coefficients $k_j(x,y)$ and the diffusivities $D_i(x,y)$ (the latter assumed to be piece-wise constant) can access tissue information such as the local cell type through a map $(x,y)\to p$. This allows for cell-specific transport and reaction kinetics. At their birth, cells can draw their own random kinetic parameters to represent cell-to-cell variability [6,7]. In interstitial space or outside of the region occupied by the tissue, background coefficients can be specified. Note that the general form of $R_i$ also allows for 0\textsuperscript{th}- or 1\textsuperscript{st}-order reactions, such as morphogen production or degradation. The coefficients $\alpha_i$ control the degree of anisotropy in the diffusion of each species.\\
    To simulate advection and dilution in moving, deforming and growing tissues, a discretized velocity field $\vec{v}(x,y)$ is interpolated on the finite-difference grid from the underlying off-lattice motion of cellular boundaries. Inside cells, inverse distance weighting (IDW) of the cell vertex velocities is used. In the extracellular space, three options are available (Supplementary Fig.~\ref{fig:1}): i) IDW within a user-defined cutoff radius, ii) bilinear interpolation or iii) zero velocity. At the lattice borders, Dirichlet and Neumann boundary conditions can be specified.\\
    For the coupling in opposite direction, all relevant cellular model parameters of PolyHoop, such as cortical tension, growth rates, etc., can be made dependent on a local readout of species concentrations $c_i$ and their gradients $\nabla c_i$, via user-specifiable C++11 lambda functions $f(p,c_1,...,c_n,\nabla c_1,...,\nabla c_n,t)$. Contact parameters (the adhesion strength and the friction coefficient) can be defined based on the properties of both cells involved in the contact, allowing to model differential adhesion for instance. In addition, a chemotactic force vector and a cell type specifier can be defined in the above functional form.\\[2ex]
	{\em Additional comments including restrictions and unusual features:} PolyMorph inherits all physical features of its predecessor PolyHoop except cell fusion, but relaxes its design paradigm of strictly minimal code somewhat, in favor of more modularity to account for the increased level of complexity. It does therefore not supersede PolyHoop, but rather represents a spin-off program. The name PolyMorph conveys the shifted focus on morphogenetic problems involving a coupling to multi-component species transport.\\
    Like PolyHoop, PolyMorph writes a series of VTK output files that can be viewed in ParaView. A structured grid file (.vts) containing user-specified lattice data is written for every animation frame. The user may further call the function \texttt{Ensemble::write\_OFF()} to save a tissue in its current state in the common Object File Format (.off), for later use as a starting point of a different simulation.\\
    Three limitations of the current implementation are that only diagonal elements in the diffusion coefficient matrix are supported, both the cell mechanics and the finite-difference solver operate with the same timestep, and both spatial directions use the same grid spacing.
    \\

\end{small}

\end{abstract}


\end{frontmatter}

\section*{CRediT authorship contribution statement}

\textbf{N.P.G.~M\"{u}ller:} Software, Validation, Visualization. \textbf{R.~Vetter:} Conceptualization, Methodology, Software, Writing - Original Draft, Supervision, Project administration.

\section*{Declaration of competing interest}

The authors declare that they have no known competing financial interests or personal relationships that could have appeared to influence the work reported in this paper.

\section*{Appendix A. Supplementary material}

\begin{figure}[!ht]
    \centering
    \includegraphics[width=0.8\linewidth]{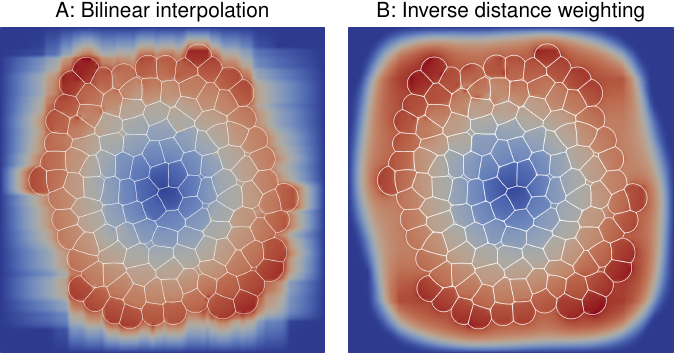}
    \caption{\textbf{Velocity field interpolation methods for the extracellular space.} Vertex velocities from the off-lattice cell boundaries (white) are used to define the velocity field on the finite-difference lattice. In the domain external to the tissue, either bilinear interpolation between the tissue boundary and the lattice boundary (A), inverse distance weighting (IDW) of all cell vertices within a user-specified radius (B), or zero velocity (not shown) can be used. IDW avoids artifacts which may be undesired in applications in which the ambient flow field matters, but is more computationally expensive. The color shows the resulting velocity magnitude (blue: 0; red: maximum).
    }
    \label{fig:1}
\end{figure}


\begin{thebibliography}{0}
		\bibitem{1}R.~Vetter, V.~Runser, D.~Iber. PolyHoop: Soft particle and tissue dynamics with topological transitions. Comput.\ Phys.\ Commun.\ 299 (2024) 109128
		\bibitem{2}K.~A.~Rejniak. A single-cell approach in modeling the dynamics of tumor microregions. Math.\ Biosci.\ Eng.\ 2 (2005) 643--655
        \bibitem{3}S.~Tanaka, D.~Sichau, D.~Iber. LBIBCell: a cell-based simulation environment for morphogenetic problems. Bioinformatics 31 (2015) 2340--2347
        \bibitem{4}B.~Merchant, L.~Edelstein-Keshet, J.~J.~Feng. A Rho-GTPase based model explains spontaneous collective migration of neural crest cell clusters. Dev.\ Biol.\ 444, Suppl.\ 1 (2018) S262--S273
        \bibitem{5}R.~Conradin, C.~Coreixas, J.~Latt, B.~Chopard. PalaCell2D: A framework for detailed tissue morphogenesis. J.\ Comput.\ Sci. 53 (2021) 101353
        \bibitem{6}R.~Vetter, D.~Iber. Precision of morphogen gradients in neural tube development. Nat.\ Commun.\ 13 (2022), 1145
        \bibitem{7}Y.~Long, R.~Vetter, D.~Iber. 2D effects enhance precision of gradient-based tissue patterning. iScience 26 (2023), 107880
	\end{thebibliography}
\end{document}